\renewcommand{\k}{{\mathbf k}}
\renewcommand{\u}{{\mathbf u}}
\newcommand{\eq}[1]{\begin{equation}#1\end{equation}}
\newcommand{\goodgap}{%
  \hspace{\subfigtopskip}%
  \hspace{\subfigbottomskip}}

\ifdefined\journal

\documentclass[acmtomacs]{acmtrans2m}
\usepackage[ruled]{algorithm2e}
\usepackage{psfig}
\usepackage{graphicx,url,subfigure,multirow,amsmath,amssymb}

\title{Graph Annotations \\ in Modeling Complex Network Topologies}

\author{
    XENOFONTAS DIMITROPOULOS \\ IBM Research, Z\"{u}rich \and
    DMITRI KRIOUKOV \\ CAIDA, University of California San Diego \and
    AMIN VAHDAT \\ University of California San Diego \and
    GEORGE RILEY \\ Georgia Tech
}

\else

\documentclass{sigcomm-alternate}
\input algorithm2e.fix
\usepackage[ruled]{algorithm2e}
\usepackage{psfig}
\usepackage{graphicx,url,subfigure,multirow,amsmath,amssymb,balance}

\begin{document}

\title{Graph Annotations \\ in Modeling Complex Network Topologies}

\numberofauthors{4}
\author{
    \alignauthor Xenofontas Dimitropoulos \\ \affaddr{IBM Research, Z\"{u}rich} \\ \email{xed@zurich.ibm.com}
    \alignauthor Dmitri Krioukov \\ \affaddr{CAIDA/UCSD} \\ \email{dima@caida.org} \and
    \alignauthor Amin Vahdat \\ \affaddr{UCSD} \\ \email{vahdat@cs.ucsd.edu}
    \alignauthor George Riley \\ \affaddr{Georgia Tech} \\ \email{riley@ece.gatech.edu}
}

\maketitle

\fi

\begin{abstract}

The coarsest approximation of the structure of a complex network, such
as the Internet, is a simple undirected unweighted graph. This
approximation, however, loses too much detail. In reality, objects
represented by vertices and edges in such a graph possess some
non-trivial internal structure that varies across and differentiates
among distinct types of links or nodes. In this work, we abstract such
additional information as network {\em annotations}. We introduce a
network topology modeling framework that treats annotations as an
extended correlation profile of a network. Assuming we have this
profile measured for a given network, we present an algorithm to
rescale it in order to construct networks of varying size that still
reproduce the original measured annotation profile.

Using this methodology, we accurately capture the network properties
essential for realistic simulations of network applications and
protocols, or any other simulations involving complex network
topologies, including modeling and simulation of network evolution. We
apply our approach to the Autonomous System (AS) topology of the Internet annotated
with business relationships between ASs. This topology captures the
large-scale structure of the Internet. In depth understanding of this
structure and tools to model it are cornerstones of research on future
Internet architectures and designs. We find that our techniques are
able to accurately capture the structure of annotation correlations within this topology,
thus reproducing a number of its important properties in synthetically-generated random graphs.

\end{abstract}

\category{C.2.1}{Network Architecture and Design}{Network topology}
\category{C.2.5}{Local and Wide-Area Networks}{Internet}
\category{G.3}{Probability and Statistics}{Distribution functions,
multivariate statistics, correlation and regression analysis}
\category{G.2.2}{Graph Theory}{Network problems}

\terms{Measurement, Design, Theory}

\keywords{Annotations, AS relationships, complex networks, topology}

\ifdefined\journal

\begin{document}

\maketitle

\fi

\section{Introduction}

Simulations of new network protocols and architectures are pointless
without realistic models of network structure and evolution.
Performance of
routing~\cite{KrKc07}, multicast~\cite{PaSt00}, and other protocols
depends crucially on network topology.  Simulations of these protocols
with inaccurate topology models can thus result in misleading
outcomes.

Inaccuracies associated with representing complex network topologies
as simple undirected unweighted graphs come not only from potential
sampling biases in topology
measurements~\cite{LaByCroXie03,ClMo05,DaAlHaBaVaVe05}, but also from
neglecting link and node {\em annotations\/}. By annotations we mean
various types of links and nodes that abstract their intrinsic
structural and functional differences to a certain degree.  For
example, consider the Internet topology at the Autonomous System (AS)
level.  Here, link annotations may represent different business
relationship between ASs, e.g., customer-to-provider, peer-to-peer,
etc.~\cite{DiKrFo06}, while node annotations may represent different
types of ASs, e.g., large or small Internet Service Providers (ISPs),
exchange points, universities, customer enterprises,
etc.~\cite{DiKrRi06}. In router-level Internet topologies, link
annotations can be different transmission speeds, latencies, packet
loss rates, etc. One can also differentiate between distinct types of
links and nodes in other networks, such as social, biological, or
transportation networks.  In many cases, simply reproducing the
structure of a given network is insufficient; we must also understand
and reproduce domain-specific annotations.

We propose network annotations as a general framework to provide the
next level of detail describing the ``microscopic'' structure of
links and nodes. Clearly, since links and nodes are constituents of
a global network, increasing description accuracy at the
``microscopic'' level will also increase overall accuracy at the
``macroscopic'' level as well.  That is, including appropriate
per-node or per-link annotations will allow us to capture and
reproduce more accurately a variety of important global graph
properties. In the AS topology case, for example, instead of
considering only shortest paths, we will be able to study the
structure of paths that respect constraints imposed by routing
policies and AS business relationships.

Higher accuracy in approximating network structure is desirable not
only for studying applications and protocols that depend on such
structure, but also for modeling network evolution. For example,
realistic Internet AS topology growth models should be based on
economic realities of the Internet since AS links are nothing but
reflections of AS contractual relationships, i.e., results of business
decisions made by organizations that the corresponding ASs
represent. Therefore, economy-based AS topology models naturally
produce links annotated with AS relationships. AS relationship
annotations are thus intrinsic to such models.

Network annotations should also be useful for researchers studying
only those networks that preserve some domain-specific constraints,
thus avoiding ``too random'' networks that violate these
constraints. Examples of such ``technological'' constraints for router
topologies include maximum node degree limits, specific relationships
between node degree and centrality, etc.~\cite{LiAlWiDo04}. In this
context, we note that any node or link attributes, including their
degrees and centrality, are forms of annotations. Therefore, one can
use the network annotation framework to introduce domain-specific or
any other constraints to work with network topologies narrowed down to
a specific class. We also note that the network annotation framework
is sufficiently general to include directed and weighted networks as
partial cases, since both link directions and weights are forms of
annotations.

After reviewing, in Section~\ref{sec:rel}, past work on network
topology modeling and generation, which largely ignores annotations,
we make the following contributions in this paper:
\begin{itemize}
\item In Section \ref{sec:short}, we demonstrate the importance of network
annotations using the specific example of AS business relationships in the
Internet.
\item In Section~\ref{sec:pro}, we introduce a general network annotation
formalism and apply it to the Internet AS topology annotated with
AS business relationships.
\item In Section~\ref{sec:mod}, we formulate a general methodology and
  specific algorithms to: i) rescale the annotation correlation
  profile of the observed AS topology to arbitrary network sizes; and
  ii) construct synthetic networks reproducing the rescaled annotation
  profiles.
While we discuss our graph rescaling and
construction techniques in the specific context of AS topologies, these
techniques are generic and can be used for generating synthetic
annotated networks that model other complex systems.
\item In Section~\ref{sec:eva}, we evaluate
the properties of the resulting synthetic AS topologies and show that
they recreate the annotation correlations observed in real annotated
AS topologies as well as other important properties
directly related to common metrics used in simulation and performance
evaluation studies.
\end{itemize}
We conclude by outlining some implications and directions for future
work in Section~\ref{sec:con}.

\section{Related work}
\label{sec:rel}

A large number of works have focused on modeling Internet topologies
and on developing realistic topology generators. Waxman~\cite{Wa88} introduced
the first topology generator that became widely known. The Waxman generator was
based on the classical (Erd\H{o}s-R\'{e}nyi) random graph model~\cite{ErRe59}.
After it became evident that
observed networks have little in common with classical random graphs,
new generators like GT-ITM~\cite{GTITM} and Tiers~\cite{tiers} tried
to mimic the perceived hierarchical network structure and were consequently called
{\it structural}.
In 1999, Faloutsos {\it et al}.~\cite{FaFaFa99} discovered
that the degree distributions of router- and
AS-level topologies of the Internet followed a power law. Structural
generators failed to reproduce the observed power laws. This failure
led to a number of subsequent works trying to resolve the problem.

The existing topology models capable of reproducing power laws can be
roughly divided into the following two classes:
causality-aware and causality-oblivious. The first class includes the
Barab\'{a}si-Albert~(BA)~\cite{AlBa00} preferential attachment model,
the Highly Optimized Tolerance~(HOT) model~\cite{CaDo99}, and their
derivatives.
The BRITE~\cite{MeLaMaBy01} topology generator belongs to this class,
as it employs preferential attachment mechanisms to generate synthetic Internet topologies.
The models in this class grow a network by
incrementally adding nodes and links to a graph based on a formalized
network evolution process. One can show that both BA and HOT growth mechanisms produce
power laws.

On the other hand, the causality-oblivious approaches try to match a
given (power-law) degree distribution without accounting for
different forces that might have driven evolution of a network to
its currently observed state. The models in this class include
random graphs with given expected~\cite{ChLu02} and
exact~\cite{AiChLu00} degree sequences, Markov graph rewiring
models~\cite{MaSneZa04,GkaMiZe03a}, and the Inet~\cite{WiJa02}
topology generator. Recent work by Mahadevan {\it et al.}\ introduced
the $dK$-series~\cite{MaKrFaVa06} extending this class of models to
account for node degree correlations of arbitrary order. Whereas
the $dK$-series provides a set of increasingly accurate
descriptions of network topologies represented as graphs, network
annotations are another, independent and ``orthogonal'' to
$dK$-series, way to provide more accurate and complete information
about actual complex systems that these graphs represent.

Frank and Strauss first formally introduced the annotated (colored)
random Markov graphs in~\cite{FS86}. In their definition, every edge
is colored by one of~$T$ colors. More recently, S\"{o}derberg
suggested a slightly different definition~\cite{Soderberg03a}, where
every half-edge, i.e., stub, is colored by one of $T$ colors. Every
edge is thus characterized by a pair of colors. This definition is
very generic. It includes uncolored and standard
colored~\cite{FS86} random graphs, random vertex-colored
graphs~\cite{Soderberg02}\footnote{Random graphs with colored nodes
are a partial case of random graphs with hidden
variables~\cite{BoPa03}.}, and random directed graphs~\cite{BoSe05}
as partial cases. S\"{o}derberg considers some analytic properties
of the ensemble of these random colored graphs
in~\cite{Soderberg03b}.
In~\cite{Soderberg05}, he observes strong
similarities between random graphs colored by $T$ colors and random
Feynman graphs representing a perturbative description of a
$T$-dimensional system from quantum or statistical mechanics.

Recent works on annotation techniques specific to AS graphs
\mbox{include~\cite{ghitle}} and~\cite{ChJaWi06}. The
\mbox{GHITLE~\cite{ghitle}} topology generator produces AS
topologies with c2p and p2p annotations based on simple design
heuristics and user-controlled parameters. The work by Chang {\it et
al}.~\cite{ChJaWi06} describes a topology evolution framework that
models ASs' decision criteria in establishing c2p and p2p
relationships. Our methodology is different in that it explores the
orthogonal, causality-oblivious approach to modeling link
annotations. Its main advantage is that it is applicable to modeling
any type of complex networks.

\section{AS relationships and why they matter}
\label{sec:short}

In this section, we introduce our specific example of network
annotations---AS relationships. We first describe what AS
relationships represent and then discuss the results of simple
simulation experiments showing why preserving AS relationship
information is important.

AS relationships are annotations of links of the Internet AS-level
topology. They represent business agreements between pairs of AS
neighbors. There are three major types of AS relationships:
1)~customer-to-provider~(c2p), connecting customer and provider ASs;
2)~peer-to-peer~(p2p), connecting two peer ASs; and
3)~sibling-to-sibling~(s2s), connecting two sibling ASs. This
classification stems from the following BGP route export policies,
dictated by business agreements between ASs:

\begin{itemize}
\item exporting routes to a provider or a peer, an
  AS advertises its local routes and routes received from its customer ASs
  only;
\item exporting routes to a customer or a sibling, an
  AS advertises all its routes, i.e., its local routes and routes received
  from all its AS neighbors.
\end{itemize}
Even though there are only two distinct export policies, they lead
to the three different AS relationship types when combined in an
asymmetric~(c2p) or symmetric~(p2p or s2s) manner.

If all ASs strictly adhere to these export policies, then one can
easily check~\cite{Gao01} that every AS path must be of the
following {\it valley-free\/} or {\it valid\/} pattern: zero or more
c2p links, followed by zero or one p2p links, followed by zero or
more p2c links, where by `p2c' links we mean c2p links in the
direction from the provider to the customer.

Routing policies reflect business agreements and economic
incentives. For this reason, they are deemed more important than
quality of service and other criteria. As a result, suboptimal
routing and inflated AS paths often occur. Gao and Wang~\cite{GaWa}
used BGP data to measure the extent of AS path inflation in the
Internet. They found that at least 45\% of the AS paths observed in
BGP data are inflated by at least one AS hop and that AS paths can
be inflated by as long as 9 AS hops.

\begin{figure}
\centerline{\ \psfig{figure=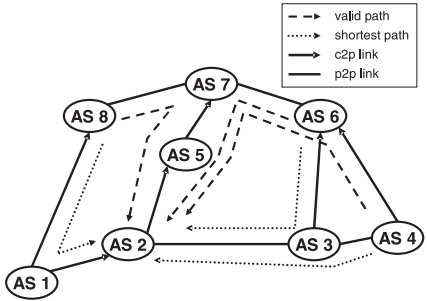,width=2.2in}\ }
\caption{
Example AS topology annotated with AS relationships. The dotted lines
represent shortest paths between ASs~4, 6 and 8 to AS~2. The dashed
lines represent policy compliant paths from the same sources to the
same destination.}
\label{figure}

\end{figure}

Taking into account such inflation effects is important for
meaningful and realistic simulation studies. For example, consider the AS
topology in Figure~\ref{figure}, which is a small part of the real
(measured) AS topology annotated with AS relationships inferred using
heuristics in~\cite{DiKrHuClRi05}.
Directed links represent c2p relationships that point towards
the provider and undirected links represent p2p relationships. If we
ignore AS relationships then the shortest paths from ASs 4, 6, and 8
to AS 2 are shown with dotted lines. On the other hand, if we account
for AS relationships these paths are no longer valid. In particular,
the path 4$\rightarrow$3$\rightarrow$2 transverses two p2p links; the
path 6$\rightarrow$3$\rightarrow$2 transverses a p2c link followed by
a p2p link; and the path 8$\rightarrow$1$\rightarrow$2 transverses a
c2p link after having gone through a p2c link. As all these paths are
not valid, they are not used in practice. The paths actually used are
the policy compliant paths marked with dashed lines.

In other words, the first effect of taking AS relationships into
account is that paths become longer than the corresponding shortest
paths. From a performance perspective, longer paths can affect metrics
such as end-to-end (e2e) delay, server response time, jitter,
convergence time, and others.

To illustrate this effect, we simulated the topology in
Figure~\ref{figure} using BGP++~\cite{DiRi03}.  We used a single
router per AS and configured appropriate export rules between ASs
according to the guidelines discussed above. We set the delay of each
link to 10 milliseconds and the bandwidth to 400kbps.  Then, we
configured exponential on/off traffic sources at ASs 4, 6 and 8 that
send traffic to AS 2 at a rate of 500kbps. We run the simulation for
120 seconds; for the first 100 seconds we waited for routers to
converge\footnote{Typically routers take much less than 100 seconds to
converge, but to be conservative we used a longer period.}  and at the
100th second we started the traffic sources. We then measured the e2e
delay between the sources and the destination with AS relationships
disabled and enabled.

\begin{figure}
\centering
\resizebox{8cm}{!}{
\includegraphics[width=3in]{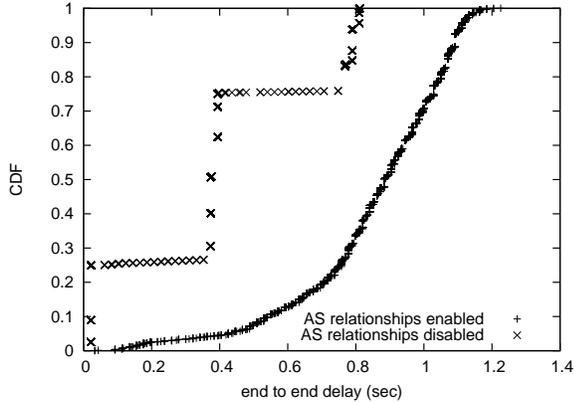}
}
\caption{CDF of e2e delay between traffic sources and destination.}
\label{e2e}
\end{figure}

In Figure~\ref{e2e} we depict the cumulative distribution function
(CDF) of the e2e delays for the both cases. We first notice that the
CDF with AS relationships enabled shifts to the
right, which means that there is a significant increase in the e2e
delay. In particular, the average e2e delay with AS relationships
enabled is~0.853 seconds, whereas without AS relationships it drops
to~0.389 seconds. Besides the decrease in the e2e delay, we see
that the CDF with AS relationships
is much smoother than the other CDF, which exhibits a step-wise
increase. The reason for that difference is that in the former
case we have more flows sharing multiple queues and, consequently,
more diverse queue dynamics, while in the latter case, almost all
paths are disjoint, leading to mostly fixed e2e delays. The observed
difference signifies that the e2e delay with AS relationships enabled
exhibits a much higher variability compared to the case with AS
relationships ignored. This difference in variability is likely
to affect other performance metrics like jitter and router buffer occupancy.

\begin{table}
\centering
\caption{Total number of paths for each AS with AS relationships
     enabled and AS relationships disabled.}
\label{table:paths}
\resizebox{8cm}{!}{
\begin{tabular}{|c|c||c|c|c|c|c|c|c|c|}
\hline
\multicolumn{2}{|c||}{AS number}  & 1 & 2 & 3 & 4 & 5 & 6 & 7 & 8 \\ \hline
\multirow{2}{*}{Number of paths} & AS relationships disabled & 12 & 13 & 16 & 15 & 13 & 15 & 15 & 13 \\
& AS relationships enabled & 12 & 9 & 10 & 8 & 8 & 7 & 9 & 6 \\
\hline\end{tabular}
}
\end{table}

Another consequence of policy-constrained routing is that ASs have
fewer alternative AS paths. For example, in Figure~\ref{figure} when
ignoring AS relationships AS~7 has three (one through each neighbor)
disjoint paths to reach destination~2. One the other hand, with AS
relationships enabled, AS~7 has only one possible path through AS~5,
since the other two paths are not valid. In Table~\ref{table:paths},
we show the total number of paths we found in the BGP tables of the
eight ASs in our simulations. The consistent decrease in the number of
paths when AS relationships are enabled highlights that ignoring AS
relationships increases the path diversity of the ASs in a
simulation. Path diversity is an important property related to network
robustness, vulnerability to attacks, links and router failures, load
balancing, multi-path routing, convergence of routing protocols, and
others.

Yet another effect of policy routing is different distribution of
load on ASs and AS links. Indeed, due to the smaller
number of available AS paths, compared to shortest path routing,
some nodes and links are likely to experience greater traffic load.
For example, in Figure~\ref{figure}
the dashed paths share the links from AS~7 to AS~5.  On the other
hand, when assuming shortest path routing the three paths are mostly
disjoint: only one link, the link between AS~3 and AS~2, is shared by two
flows. Thus, AS links and nodes will receive greater load, compared to
the case with AS relationships ignored. Higher load is likely to
produce more packet loss, increased delay, congestion, router
failures, and other undesirable effects.
In Table~\ref{table:bw} we list the average bandwidth in our
simulations for each of the three flows with and without AS
relationships enabled. We find that because of the increased
load on the links between AS~7 and AS~5 the average bandwidth
of the three flows decreases substantially.

\begin{table}
\centering
\caption{Average bandwidth per flow with AS relationships enabled or disabled.}
\label{table:bw}

\resizebox{8cm}{!}{
\begin{tabular}{|c|c||c|c|c|}
\hline
\multicolumn{2}{|c||}{Flow}  & 4 $\rightarrow$ 2 & 6 $\rightarrow$ 2 & 8 $\rightarrow$ 2 \\ \hline
\multirow{2}{*}{Bandwidth (Kbps)} & AS relationships disabled & 202 & 196 & 397 \\
& AS relationships enabled & 113 & 164 & 121 \\ \hline
\end{tabular}
}

\end{table}

To summarize this section, we have provided three
examples showing that ignoring AS relationship annotations leads to
inaccuracies, which make
the corresponding properties look ``better'' than they are in
reality. Indeed, if AS relationships are ignored, then:
\begin{itemize}
\item paths are shorter than in reality;
\item path diversity is larger than in reality; and
\item traffic load is lower than in reality.
\end{itemize}

\section{Network topology annotations} \label{sec:pro}

In this section we first introduce our general formalism to annotate
network topologies. We then show how this formalism applies to our
example of the AS-level Internet topology annotated with AS
relationships.

\subsection{General formalism} \label{sec:general-formalism}

Our general formalism is close to random colored graph definitions
from~\cite{Soderberg03a} and borrows parts of the convenient
$dK$-series terminology from~\cite{MaKrFaVa06}.

We define the annotated network as a graph $G(V,E)$, \mbox{$|V|=n$}
and \mbox{$|E|=m$}, such that all $2m$~edge-ends (stubs) of all
$m$~edges in~$E$ are of one of several colors~$c$, \mbox{$c=1 \ldots
C$}, where $C$ is the total number of stub colors. We also allow for
node annotations by an independent set of node colors
\mbox{$\theta=1 \ldots \Theta$}. We do not use node annotations in
this paper and we do not include them in the expressions below in
order to keep them clearer. It is however trivial to add node
annotations to these expressions.

\begin{figure}
\centering
\includegraphics[width=3in]{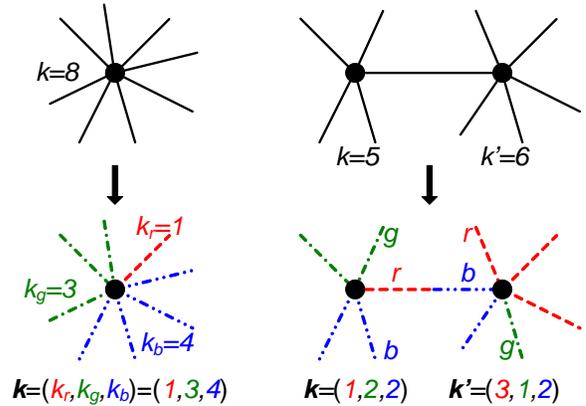}
\caption{
 {\bf The $1K$- and $2K$-annotations.} Three different stub colors are
 represented by dashed (color {\it red}), dash-dotted (color {\it
 green}), and dash-double-dotted (color {\it blue}) lines.
}
\label{fig:annotations}
\end{figure}

Compared to the non-annotated case when the node degree is fully
specified by an integer value~$k$ of the number of stubs attached to
the node, we now have to list the numbers of attached stubs of each
color to fully describe the node degree. Instead of scalar~$k$, we
thus have the node degree vector
\[\k=(k_1,\ldots,k_C),\]
which has $C$ components~$k_c$, each specifying the number of
$c$-colored stubs attached to the node, cf.~the left side of
Figure~\ref{fig:annotations}. The $L^1$-norm of this vector yields the
node degree with annotations ignored,
\eq{k = |\k|_1 = \sum_{c=1}^C k_c. \label{eq:scalar-k}}

The number~$n(\k)$ of nodes of degree~$\k$ defines the node degree
distribution,
\begin{equation}\label{eq:1k-annotated}
\frac{n(\k)}{n}\xrightarrow[n\to\infty]{}P(\k),
\end{equation}
in the large-graph limit. We can think of~$n(\k)$ as a
non-normalized form of~$P(\k)$. From the statistical perspective,
the $n(\k)$ ($P(\k)$) distribution is a multivariate distribution.
Its C marginal distributions are the distributions of node degrees
of each color~$c$:
\eq{n(k_c) = \sum_{\k' \; \big| \; k'_c = k_c} n(\k'),
\label{eq:marginal-1k}}
where the summation is over all vectors~$\k'$ such that their $c$'s
component is equal to~$k_c$. The degree distribution~$n(\k)$ thus
represents per-node correlations of degrees of different colors.
Following the terminology in~\cite{MaKrFaVa06}, we call the node
degree distribution $n(\k)$ ($P(\k)$) the {\em $1K$-annotated
distribution}.

We then define the {\em $2K$-annotated distribution\/} as
correlations of annotated degrees of connected nodes, or simply as
the number of edges that have stub of color~$c$ connected to a node
of degree~$\k$ and the other stub of color~$c'$ connected to a node
of degree~$\k'$, $n(c,\k;c',\k')$. See the right side of
Figure~\ref{fig:annotations} for illustration.

As in the non-annotated case, the $2K$-distribution yields a more
exhaustive statistics about the annotated network topology and fully
defines the $1K$-distribution. To see that, we introduce the
following notations:
\begin{eqnarray*}
    \tilde{\k} &=& (c,\k),\\
    \tilde{\k}' &=& (c',\k'),\\
    \mu(c,c') &=& 1 + \delta(c,c'),\\
    \mu(\k,\k') &=& 1 + \delta(\k,\k'),\\
    \mu(\tilde{\k},\tilde{\k}') &=& 1 +
    \delta(\tilde{\k},\tilde{\k}'),
\end{eqnarray*}
where~$\delta(x,x')$ is the standard Kronecker delta:
\[
    \delta(x,x') =
        \begin{cases}
        1 & \text{if \; $x=x'$,}\\
        0 & \text{otherwise,}
        \end{cases}
\]
and~$x$ is either~$c$, $\k$, or $\tilde{\k}$. With these notations,
one can easily check that the normalized $2K$-annotated distribution
is
\eq{ P(\tilde{\k},\tilde{\k}') = n(\tilde{\k},\tilde{\k}')
\mu(\tilde{\k},\tilde{\k}') / (2m),
\label{eq:2k-annotated} }
the number of edges of any pair of colors connecting nodes of
degrees $\k$ and $\k'$ is
\[
n(\k,\k') = \sum_{c,c'} n(\tilde{\k},\tilde{\k}')
\mu(\tilde{\k},\tilde{\k}') / \mu(\k,\k'),
\]
the normalized form of this distributions is
\[ P(\k,\k') = n(\k,\k') \mu(\k,\k') / (2m),
\]
and the $1K$-distribution is given by
\begin{eqnarray}
n(\k) &=& \sum_{\k'} n(\k,\k') \mu(\k,\k') / k, \label{eq:1k-from-2k-n}\\
P(\k) &=& \frac{\bar{k}}{k} \sum_{\k'} P(\k,\k'), \label{eq:1k-from-2k-p}
\end{eqnarray}
where $\bar{k}=2m/n$ is the average degree. The last two expressions
show how one can find the $1K$-annotated distribution given the
$2K$-annotated distribution, and they look exactly the same as in
the non-annotated case~\cite{MaKrFaVa06}, except that we have
vectors~$\k,\k'$ instead of scalars~$k,k'$.

The $dK$-annotated distributions with $d>2$~\cite{MaKrFaVa06} can be
defined in a similar way.

\subsection{The AS relationship annotations}
\label{sec:as-annotations}

In the specific case of the AS-level Internet topology that
interests us in this paper, we have just three colors: customer,
provider, and peer.\footnote{We ignore sibling relationships, since
they typically account for a very small fraction of the total number
of edges. As found in~\cite{DiKrFo06}, the number of s2s edges is
only 0.46\% of the total number of edges in the AS-level Internet.}
We assign the following numeric values to represent these three
colors:
\[ c =
\begin{cases}
1 & \; \text{customer},\\
2 & \; \text{provider},\\
3 & \; \text{peer}.
\end{cases}
\]
These three stub annotations come under the following two
constraints defining the only two types of edges that we have:
1)~{\em c2p edges}: if one stub of an edge is customer, then the
other stub of the same edge is provider, and vice versa; and 2)~{\em
p2p edges}: if one stub of an edge is peer, then the other stub of
the same edges is also peer. The c2p edges are thus asymmetric,
i.e., a generalization of directed edges, while the p2p edges are
symmetric, i.e., a generalization of bi-directed or undirected
edges.

While the $2K$-annotated distribution~$n(\tilde{\k},\tilde{\k}')$
contains the most exhaustive information about the network topology,
it has too many (seven) independent arguments. As a result, the full
$2K$-annotated statistics is extremely sparse, which makes it
difficult to model and reproduce directly. We thus have to find some
summary statistics of~$n(\tilde{\k},\tilde{\k}')$ that we can model in
practice. For each concrete complex network type, these summary
statistics might be different. Given measurement data for a specific
complex network, one would usually have to start with identifying a
meaningful set of summary statistics of the $2K$-annotated
distribution, and then proceed from there. At the same time, we
believe that as soon as the $2K$-annotated distribution fully defines
an observed complex network, i.e., the network is
$2K$-annotated-random~\cite{MaKrFaVa06}, one can generally use the set
of summary statistics that we found necessary and sufficient to
reproduce in order to model correctly the Internet AS topology.  In
the rest of this section, we list these statistics and describe the
specific meanings that they have in the AS topology case.

{\bf Degree distribution (DD).} This statistics is the traditional
non-annotated degree distribution~$n(k)$, where $k$ is as in
eq.~(\ref{eq:scalar-k}). The DD tells us how many ASs of each total
degree~$k$ are in the network.

{\bf Annotation distributions (ADs).} The DD of an AS topology does
not convey any information about the AS relationships. The initial
step to account for this information is to reproduce the distributions
of ASs with specific numbers of attached customer, provider, or peer
stubs. These annotation distributions~(ADs) are the marginal
distributions~$n(k_c)$, $c=1,2,3$, of the $1K$-annotated
distribution. They are given by eq.~(\ref{eq:marginal-1k}).
If $k_1$ ($k_2$) customer (provider) stubs attach to an AS, then
this AS has exactly $k_1$ ($k_2$) providers (customers), since the
c2p edges are asymmetric. Consequently, the ADs~$n(k_1)$ and
$n(k_2)$ tell us how many ASs with the specific numbers of providers
and, respectively, customers the network has. Since the p2p edges
are symmetric, the AD~$n(k_3)$ is the distribution of ASs with
specific numbers of peers.

{\bf Annotated degree distribution (ADD).} The ADs do not tell us
anything about the correlations among annotated degrees of the same
node, i.e., how many customers, providers, {\em and},
simultaneously, peers a specific AS has. Correlations of this type
are fully described by the $1K$-annotated distribution in
eq.~(\ref{eq:1k-annotated}), which we also call the annotated degree
distribution~(ADD). These correlations  are present in the Internet.
For example, large tier-1 ISPs typically have a large number of
customers, i.e., large~$k_2$, no providers, i.e., zero~$k_1$, and a
small number of peers, i.e., small~$k_3$. On the other hand,
medium-size ISPs tend to have a small set of customers, several
peers, and few providers.
Ignoring the ADD while generating synthetic graphs can lead to
artifacts like high-degree nodes with many providers---a property
obviously absent in the real Internet.

{\bf Joint degree distributions (JDDs).} While the ADD contains the
full information about degree correlations ``at nodes,'' it does not
tell us anything about degree correlations ``across links,'' while
the latter type of correlations is also characteristic for the
Internet. For example, large tier-1 ISPs typically have p2p
relationships with other tier-1 ISPs, not with much smaller ISPs,
while small ISPs have p2p links with other small ISPs. In other
words, p2p links usually connect ASs of similar degrees, i.e.,
\mbox{$k \sim k'$}. Similarly, c2p links tend to connect low-degree
customers to high-degree providers, i.e., \mbox{$k \ll k'$}. If we
ignore these correlations, we can synthesize graphs with
inaccuracies like p2p links connecting ASs of drastically dissimilar
degrees. To reproduce these correlations, we work with the following
summary statistics of the $2K$-annotated distribution in
eq.~(\ref{eq:2k-annotated}):
\begin{eqnarray}
n_{c2p}(k,k') &=& \sum_{\k,\k' \; \big| \; |\k|_1 = k , \, |\k'|_1 = k'}
    n(1,\k;2,\k'), \label{eq:jdd-c2p}\\
n_{p2p}(k,k') &=& \sum_{\k,\k' \; \big| \; |\k|_1 = k , \, |\k'|_1 = k'}
    n(3,\k;3,\k'), \label{eq:jdd-p2p}
\end{eqnarray}
where the summation is over such vectors $\k$ and $\k'$ that their
$L^1$-norms are $k$ and $k'$ respectively. The first expression
gives the number of c2p links that have their customer stub attached
to a node of total degree~$k$ and provider stub attached to a node
of total degree~$k'$. The second expression is the number of p2p
links between nodes of total degrees $k$ and $k'$. In other words,
these two objects are the joint degree distributions (JDDs) for the
c2p and p2p links.

In summary, we work with the four types of distributions, i.e., DD,
ADs, ADD, and JDDs, that allow two types of classification:
\begin{enumerate}
\item {\bf Univariate vs.\ multivariate distributions:}
\begin{enumerate}
\item \underline{Univariate.}
The ADs and DD are distribution of only one random variable.
\item \underline{Multivariate.}
The ADD and JDDs are joint distribution of three and two random
variables. The marginal distribution of these variables are the ADs,
cf.~eq.~(\ref{eq:marginal-1k}), and DD,
cf.~eqs.~(\ref{eq:jdd-c2p},\ref{eq:jdd-p2p}), respectively.
\end{enumerate}
\item {\bf $\mathbf{1K}$- vs.\ $\mathbf{2K}$-summary statistics:}
\begin{enumerate}
\item \underline{$1K$-derived.}
The DD, ADs, and ADD are fully defined by the $1K$-annotated
distribution: that is, we do not need to know the $2K$-annotated
distribution to calculate the distributions in this class.
\item \underline{$2K$-derived.}
The JDDs are fully defined only by the $2K$-annotated distribution.
Note that it also defines the $1K$-annotated distribution via
eqs.~(\ref{eq:1k-from-2k-n},\ref{eq:1k-from-2k-p}).
\end{enumerate}
\end{enumerate}

\section{Generating annotated AS graphs} \label{sec:mod}

In this section we describe how we generate synthetic annotated AS
graphs of arbitrary sizes. We want our synthetic graphs to reproduce
as many important properties of the original measured topology as
possible. For this purpose we decide to explicitly model and
reproduce the summary statistics of the $2K$-annotated distribution
from Section~\ref{sec:as-annotations}, because \cite{MaKrFaVa06}
showed that by reproducing $2K$-distributions, one automatically
captures a long list of other important properties of AS topologies.
In other words, the task of generating synthetic annotated
topologies becomes equivalent to the task of generating random
annotated graphs that reproduce the summary statistics of the
$2K$-annotated distribution of the measured AS topology.

We wish to be able to generate synthetic topologies of different
sizes, but the $2K$-summary statistics defined in
Section~\ref{sec:as-annotations} are all bound to a specific graph
size. Therefore, in order to generate arbitrarily-sized graphs, we
need first to rescale the $2K$-summary statistics from the original
to target graph sizes.
We say that an empirical distribution is {\em rescaled\/} with
respect to another empirical distribution, if the both distributions
are defined by two different finite collections of random numbers
drawn from the same continuous distribution. For example, the
distributions of node scalar (or vector) degrees in two different
graphs are rescaled with respect to each other if these degrees are
drawn from the same continuous univariate (or multivariate)
probability distribution.
We say that a $2K$-annotated graph is {\em rescaled\/} with respect
to another $2K$-annotated graph, if all the $2K$-summary statistics
of the first graphs are rescaled with respect to the corresponding
$2K$-summary statistics of the second graph. This definition of
rescaling is equivalent to assuming that for each summary statistic,
the same distribution function describes the ensemble of empirical
distributions of the statistic in past, present, and future Internet
topologies. In other words, we assume that the $2K$-annotated
correlation profile of the Internet AS topology is an invariant of
its evolution. This assumption is realistic, as discussed, for example,
in \cite{PaSaVe04-book},
where it is shown that the
non-annotated $1K$- and $2K$-distributions of the Internet AS topology
have stayed approximately the same during all the years (more than a
decade) of the existing data time span.

\begin{figure*}
\centering

\subfigure[Original distribution of 19,036 peer-degrees in the measured AS topology.]{\label{fig:ex:1}\includegraphics[width=.35\textwidth]{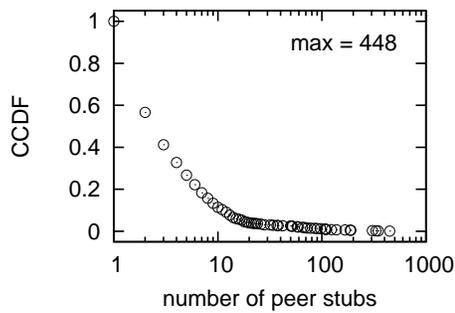}}\goodgap
\subfigure[Continuous distribution approximating the original one.]{\label{fig:ex:2}\includegraphics[width=.35\textwidth]{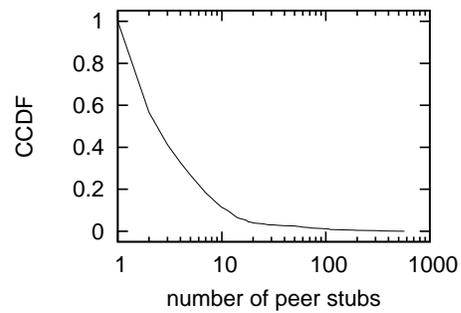}}\\
\subfigure[Rescaled distribution of 5,000 samples.]{\label{fig:ex:3}\includegraphics[width=.31\textwidth]{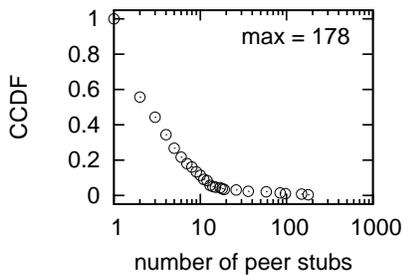}}\goodgap
\subfigure[Rescaled distribution of 20,000 samples.]{\label{fig:ex:4}\includegraphics[width=.31\textwidth]{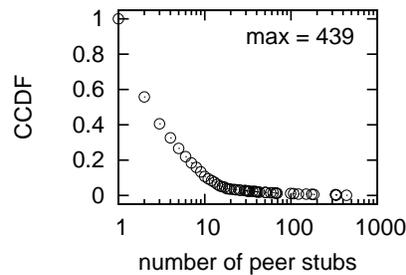}}\goodgap
\subfigure[Rescaled distribution of 50,000 samples.]{\label{fig:ex:5}\includegraphics[width=.31\textwidth]{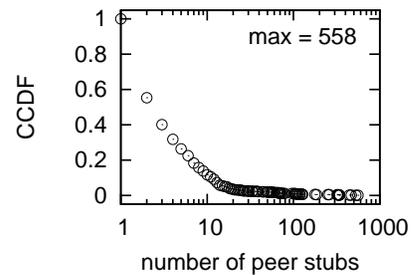}}%

\caption{\label{fig:ex} Rescaling an empirical distribution. The
distributions in the three bottom figures are rescaled with respect
to the empirical distribution in the first figure. The distribution
in the second figure is a continuous approximation of the
distribution in the first figure and is used to generate the
rescaled distributions in the bottom figures. For each discrete
distribution, we show its maximum in the top-right corners of the
plots.}

\end{figure*}

To illustrate what we mean by {\em rescaling}, consider the
empirical distribution of peer degrees, i.e., the AD $n(k_3)$, in
the measured AS topology annotated with AS relationships in
Figure~\ref{fig:ex:1}. The figure shows the empirical complementary
cumulative distribution function (CCDF) for peer-degrees of 19,036
nodes, i.e., 19,036 numbers of peer stubs attached to a node, and
the largest such number is 448. The continuous probability
distribution of Figure~\ref{fig:ex:2} approximates the empirical
distribution in Figure~\ref{fig:ex:1}.
Figures~\ref{fig:ex:3}, \ref{fig:ex:4}, and~\ref{fig:ex:5} show the
CCDFs of three collections of 5,000, 20,000, and 50,000 random
numbers drawn from the probability distribution in
Figure~\ref{fig:ex:2}.
According to our definition of rescaling, the distributions in
Figures~\ref{fig:ex:3}, \ref{fig:ex:4}, and~\ref{fig:ex:5} are
rescaled with respect to the distribution of Figure~\ref{fig:ex:1}.
We see that all the empirical distributions have the same overall
shape, but differ in the total number of samples and in the maximum
values within these sample collections. Distributions with larger
maximums correspond, as expected, to bigger collections of samples.

\subsection{Overview of the approach}
\label{sec:mod:over}

We now move to describing the details of our approach, which
consists of the following three major phases:
\begin{enumerate}

\item {\bf Extraction.}

We first extract the empirical $2K$-summary distributions from
available AS topology measurement data.
We annotate links of
the AS topology extracted from this data using existing AS
relationship inference heuristics. This extraction step is
conceptually simplest. On its output, we obtain the extracted
$2K$-summary distributions that are all bound to the size of the
measured AS graph.

\item {\bf Rescaling.}
\begin{enumerate}
\item We use the extracted empirical distributions to find their
continuous approximations.
Referring to our example in Figure~\ref{fig:ex}, this step
corresponds to computing the continuous probability distribution in
Figure~\ref{fig:ex:2} based on the empirical distribution in
Figure~\ref{fig:ex:1}.

\item We then use the computed probability distributions to rescale
the empirical distributions obtained at the extraction step. We
generate a desired, target number of random scalar or vector degree
samples drawn from the corresponding probability distributions. The
generated degree samples have empirical distributions that are
rescaled with respect to the corresponding empirical distributions
of the measured topology. Referring to our example in
Figure~\ref{fig:ex}, this step corresponds to generating the
rescaled empirical distributions in Figures~\ref{fig:ex:3},
\ref{fig:ex:4}, and~\ref{fig:ex:5} based on the probability
distribution in Figure~\ref{fig:ex:2}.
\end{enumerate}
\item {\bf Construction.}

Finally, we develop algorithms to generate synthetic graphs that
have their $2K$-summary distributions equal to given distributions,
i.e., to the corresponding distributions obtained at the previous
step. The generated graphs thus reproduce the rescaled replicas of
the $2K$-annotated distribution of the original topology, but they
are ``maximally random'' in all other respects.

\end{enumerate}

In the rest of this section, we describe each of these phases in
detail.

\subsection{Extraction}
\label{sec:extraction}

We extract the AS topology from the RouteViews~\cite{routeviews}
data, performing some standard data cleaning, such as ignoring
private AS numbers, AS sets, etc.~\cite{MaKrFo06}
The resulting AS graph is initially non-annotated. To annotate it,
we infer c2p and p2p relationships for AS links using the heuristics
in~\cite{DiKrFo06}. We thus obtain the real Internet AS topology
annotated with c2p and p2p relationships. Given this annotated
topology, we straightforwardly calculate all the empirical
$2K$-summary distributions that we have defined in
Section~\ref{sec:as-annotations}.

While the extraction phase is conceptually and technically the
simplest phase of the overall approach, it is its basis. Therefore
the quality of the input Internet topology data is a natural
concern. This data is known to exhibit a variety of vagaries, e.g.,
due to sampling biases~\cite{LaByCroXie03,ClMo05,DaAlHaBaVaVe05}.
However, our approach is oblivious with respect to data quality. It
takes any available data, extracts the described statistics from it,
and reproduces them, properly rescaled, in random synthetic graphs.
A given input topology data set thus defines an ensemble of random
graphs generated by our method. By construction, all graphs in this
ensemble reproduce the described set of annotated distributions. In
addition, in Section~\ref{sec:eva}, we perform sensitivity analysis
in order to see the strength of fluctuations of these and other
basic graph metrics within an ensemble. The quality of these graph
ensembles, in terms of how veraciously they reflect reality, will
improve as the quality of available topology data improves in the
future. In this paper, we simply illustrate our approach with the
currently available topology data. The RouteViews~\cite{routeviews}
is just one of very few sources of such data~\cite{MaKrFo06}. We
select it because it appears to be the most frequently cited
Internet topology data source.

\subsection{Rescaling}
\label{sec:rescale}

Our rescaling approach differs for univariate and multivariate
distributions.

\subsubsection{Rescaling univariate distributions}
\label{sec:rescaling-univariate}

We recall from the end of Section~\ref{sec:as-annotations} that we
have the following two types of univariate distributions: the ADs
and the DD. Here we describe how we rescale ADs. We note that we do
not have to rescale the DD the same way. The reason is that our
approach to rescaling the ADD, which we discuss below in
Section~\ref{sec:rescaling-multivariate}, automatically takes care
of rescaling the DD, since the ADD is the distribution of degree
vectors and the DD is the distribution of the $L^1$-norms of these
vectors, cf.~eq.~(\ref{eq:scalar-k}).

The first problem we face trying to compute a continuous
approximation for a given finite empirical distribution is that we
have to not only interpolate between points of the empirical
distribution, but also extrapolate above its maximum value. For
example, if we want to construct a synthetic graph bigger than the
original, then we expect its maximum degree to be larger than the
maximum degree in the original graph. Therefore we have to properly
extrapolate the observed degree distribution beyond the observed
maximum degree.

We solve this problem by fitting the univariate empirical
distributions with {\em smoothing splines}. Spline smoothing is a
non-parametric estimator of an unknown function represented by a
collection of empirical data points. Spline smoothing produces a
smooth curve passing through or near the data points. For example,
the curve in Figure~\ref{fig:ex:2} is a smooth spline of the
empirical distribution of Figure~\ref{fig:ex:1}. Spline smoothing
can also extrapolate the shape of an empirical function beyond the
original data range.

Another reason to select spline smoothing is that it comes useful
for fitting distributions that do not closely follow regular shapes,
e.g., ``clean'' power laws. The ADs of the Internet topology do not
necessarily have such regular shapes.
For example, the distribution of the number of peers, i.e., the AD
$n(k_3)$, has a complex shape that we found impossible to fit with
any single-parametric distribution.

Among available implementations of spline smoothing techniques, we
select the one in the {\tt smooth.spline} method of the R
project~\cite{r-project}, a popular statistical computing package.
The specific details of this technique are in \cite{ChHa92}.

We can approximate with splines
either the CDFs or CCDFs\footnote{Recall that the CCDF of CDF $F(x)$
is \mbox{$1-F(x)$}.} of the ADs obtained at the extraction step.
We chose to fit the CCDFs rather than the CDFs because the former
better capture the shapes of high-degree tails of our heavy-tailed
ADs.

Another important detail is that we can define an empirical CCDF to
be either a left- or right-continuous step function~\cite{stepfun}.
Usually an empirical CCDF at some point~$x$ is defined as the
fraction of samples with values strictly larger than~$x$, which
means that the distribution is right-continuous and that the
probability of a value larger than the observed maximum value is
zero, whereas the probability of a value smaller than the observed
minimum value is unspecified. For degree distributions, we know that
the probability of a degree smaller than zero is zero,\footnote{For
a given node, some but not all the degrees~$k_1$, $k_2$, and~$k_3$
can be equal to zero.} but we do not know the probability of a
degree larger than the maximum observed degree. For this reason, we
decide to fit the left-continuous variants of empirical CCDFs, i.e.,
we define a CCDF at some point~$x$ as the fraction of samples with
values greater than or equal to~$x$.

Having the original ADs fitted with splines and assuming that our
target graph size is~$N$, we finally use the standard, inverse-CDF
method to produce~$N$ random numbers that follow the continuous
distributions given by the splines. Recall that the inverse-CDF
method is based on the observation that if the CDF of $N$ random
numbers $x_j$, \mbox{$j=1 \ldots N$}, closely follows some function
$F(x)$, then the distribution of numbers \mbox{$y_j = F(x_j)$} is
approximately uniform in the interval~\mbox{$[0,1]$}. As its name
suggests, the inverse-CDF inverts this observation and operates as
follows \cite{HoLe04-book}: given a target CDF $F(x)$ and a target
size $N$ of a collection of random samples, the method first
generates $N$ random numbers $y_j$ uniformly distributed in~$[0,1]$
and then outputs numbers $x_j=F^{-1}(y_j)$, where $F^{-1}(y)$ is the
inverse of CDF $F(x)$, i.e., \mbox{$F^{-1}(F(x))=x$}. The CDF of
numbers $x_j$ closely follows $F(x)$. Figures~\ref{fig:ex:3},
\ref{fig:ex:4}, and~\ref{fig:ex:5} show random numbers generated
this way. These numbers follow the distribution in
Figure~\ref{fig:ex:2}.
To compute values of inverse CDFs on $N$ random numbers uniformly
distributed in $[0,1]$ in our case, we use the {\tt
  predict.smooth.spline} method of the R project.
Since the random numbers produced in this way are not, in general,
integers, we convert them to integer degree values using the floor
function. We have to use the floor and not the ceiling function
because we work with left- rather than right-continuous
distributions.

The outcome of the described process is three sets of $N$ random
numbers that represent $N$~customer degrees~$d^j_1$, $N$~provider
degrees~$d^j_2$, and $N$~peer degrees~$d^j_3$ of nodes in the target
graph, \mbox{$j=1 \ldots N$}. We denote the CDFs of these random
numbers by $D_1(d_1)$, $D_2(d_2)$, and $D_3(d_3)$ respectively. By
construction, these distributions are properly rescaled versions of
the customer-, provider- and peer-annotation distributions~(ADs) in
the measured AS topology.

\subsubsection{Rescaling multivariate distributions}
\label{sec:rescaling-multivariate}

Rescaling multivariate distributions is not as simple as rescaling
univariate distributions. Our approach for rescaling univariate
distributions is not practically applicable to rescaling
multivariate distributions because it is difficult to fit
distributions that have many variates and complex shapes. To rescale
multivariate distributions, we use {\em copulas}~\cite{Copulas},
which are a statistical tool for quantifying correlations between
several random variables. Compared to other well-known correlation
metrics, such as Pearson's coefficient, copulas give not a single
scalar value but a function of several arguments that fully
describes complex, fine-grained details of the structure of
correlations among the variables, i.e., their correlation profile.

According to Sklar's theorem~\cite{sklar}, any $p$-dimensional
multivariate CDF~$F$ of $p$ random variables~\mbox{$\k =
(k_1,\ldots,k_p)$} can be written in the following form:
\begin{equation}\label{eq:copula}
F(\k) = H(\u),
\end{equation}
where $\u$ is the $p$-dimensional vector composed of the $F$'s marginal CDFs
$F_m(k_m)$, \mbox{$m=1,\ldots,p$}:
\begin{eqnarray}
F_m(k_m) &=& F(\infty, \ldots, \infty, k_m, \infty, \ldots, \infty),\\
\u &=& (F_1(k_1),\ldots,F_p(k_p)).\label{eq:u}
\end{eqnarray}
The function~$H$ is called a copula and each of its marginal
distributions is uniform in~$[0,1]$.

Copulas play a critical role at the following two steps in our
approach for rescaling multivariate distributions. First, they allow
us to split a multivariate distribution of the original, measured
topology into two parts: the first part consists of the marginal
distributions~$F_m$, while the second part is their correlation
profile, i.e., copula~$H$. These two parts are independent.
Therefore, we can {\em independently\/} rescale the marginal
distributions and the correlation profile. This property
tremendously simplifies the rescaling process. The marginals are
univariate distributions that we rescale as in
Section~\ref{sec:rescaling-univariate}, while this section contains
the details of how we rescale the correlation profile. We use
copulas the second time to merge together rescaled marginals and
their correlation profile to yield a rescaled multivariate
distribution in its final form.

\begin{figure*}[t]
\centering
\includegraphics[width=5in,keepaspectratio,clip]{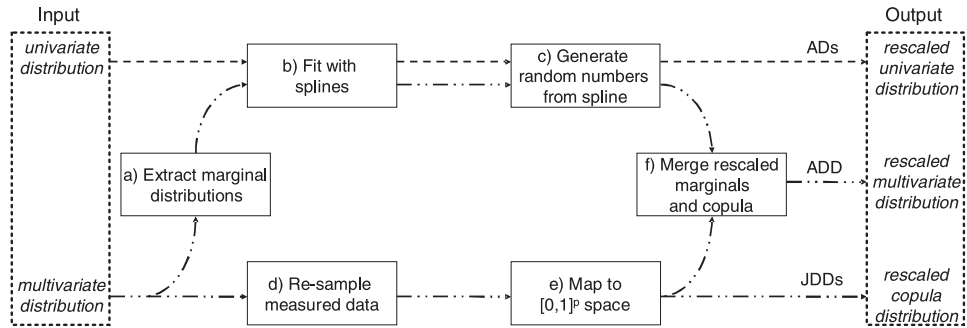}
\caption{\label{fig:road} Overview of rescaling univariate and
multivariate distributions.}

\end{figure*}

In Figure~\ref{fig:road} we present a high-level overview of our
approaches for rescaling univariate and multivariate distributions.
To rescale an original empirical univariate distribution, we first
approximate it with splines and then use these splines to generate
random numbers. We split the process of rescaling an original
empirical multivariate distribution into two independent rescaling
sub-processes, i.e., rescaling the marginals and their copula. We
rescale the marginals as any other univariate distributions. To
rescale the copula, we re-sample measured correlation data as we
describe below in this section. At the end of multivariate
rescaling, we merge the rescaled marginals with the rescaled copula
to yield a rescaled multivariate distribution. One can see from
Figure~\ref{fig:road} that multivariate rescaling is a ``superset,''
in terms of actions involved, of univariate rescaling. The following
three steps summarize the high-level description of our multivariate
rescaling approach:
\begin{enumerate}

\item extract and rescale the univariate marginals of a multivariate
distribution as described in Section~\ref{sec:rescaling-univariate}
(boxes~(a), (b), and~(c) in Figure~\ref{fig:road});

\item extract and rescale the copula of the multivariate distribution
(boxes~(d) and~(e) in Figure~\ref{fig:road}); and

\item merge the rescaled marginals and copula yielding a rescaled
multivariate distribution (box~(f) in Figure~\ref{fig:road}).

\end{enumerate}
In the rest of this section, we provide the low-level details for
the last two steps, using the ADD multivariate distribution as an
example.

At Step~2, we compute a rescaled ADD copula as follows. The
collected AS topology has $n$ nodes, and for each node~$i$,
\mbox{$i=1 \dots n$}, we record its degree
vector~$\k_i=(k^i_1,k^i_2,k^i_3)$ producing an $n$-sized set of
degree triplets. We then perform statistical simulation on this set
to produce another set of a desired size that has the same
correlations as the measured data~$\k_i$. Specifically, we
re-sample, uniformly at random and with replacement, $N$ degree
triplets from the set of vectors~$\k_i$, where $N$ is the target
size of our synthetic topology. We thus obtain an $N$-sized set of
random triplets~$\k_j$, \mbox{$j=1 \ldots N$}, and we denote their
joint CDF by~$F(\k)$. By construction, the empirical distribution of
triplets~$\k_j$ has the same correlation profile as original
triplets~$\k_i$. This procedure corresponds for box~(d) in
Figure~\ref{fig:road}.

Next, see box~(e) in Figure~\ref{fig:road}, we compute the empirical
copula of distribution~$F(\k)$. By definition, the copula of~$F(\k)$
is simply the joint distribution of vectors~$\u$ in
eqs.~(\ref{eq:copula},\ref{eq:u}). Therefore, we first compute the
marginal CDFs $F_1(k_1)$, $F_2(k_2)$, and $F_3(k_3)$ as CDFs of the
first, second, and third components of vectors~$\k_j$:
\begin{equation}
u_m^j = F_m(k_m^j) = r_m^j/N, \quad m = 1,2,3,
\end{equation}
where~$r_m^j$ is the rank (position number) of value~$k_m^j$ in the
$N$-sized list of values~$k_m$ sorted in the non-decreasing order.
Random triplets $\u_j = (F_1(k_1^j),F_2(k_2^j),F_3(k_3^j))$ are
uniformly distributed in the cube $[0,1]^3$, and their joint
CDF~$H(\u)$, $\u = (F_1(k_1),F_2(k_2),F_3(k_3))$, is the empirical
copula for distribution~$F(\k)$, cf.~eq.~(\ref{eq:copula}), that
describes the correlations among~$k_1$, $k_2$, and $k_3$.

At Step 3, box~(f) in Figure~\ref{fig:road}, we merge the rescaled
marginals~$D_m(d_m)$, $m=1,2,3$, from
Section~\ref{sec:rescaling-univariate} and copula~$H(\u)$ by
computing the target graph degree triplets $\mathbf{q}_j =
(q_1^j,q_2^j,q_3^j)$, \mbox{$j=1,\ldots,N$}, as
\begin{equation}
q_m^j = D_m^{-1}(u_m^j),
\end{equation}
where~$D_m^{-1}$ are inverse CDFs of~$D_m$ from
Section~\ref{sec:rescaling-univariate}.  By construction, the
correlation profile of annotation-degree vectors~$\mathbf{q}_j$ is
the same as of the ADD in the original topology, while the
distributions of their components~$q_m^j$ are rescaled ADs.

Algorithm~\ref{rescale} lists the described low-level details of our
multivariate rescaling, using the ADD as an example.
\begin{algorithm}[th]

\caption{Rescaling ADD}\label{rescale} \dontprintsemicolon \SetLine

\KwIn{Degree vectors $\k_i = (k_1^i,k_2^i,k_3^i)$,
    \mbox{$i=1 \dots n$}, of the measured topology;}
\KwIn{Size~$N$ of the target synthetic topology.}

\BlankLine \tcp{Step 1: AD rescaling} \ForAll{$m=1,2,3$} {
Let $k_m$ be the list of the $m^\text{th}$ component values of vectors~$\k_i$;\\
Approximate distribution $k_m$ by a smoothing spline $S_m$;\\
Sample $N$ numbers $d_m^j$, \mbox{$j=1 \ldots N$},
with probability distribution given by~$S_m$;\\
Let $D_m(d_m)$ be the CDF of $d_m$. }

\BlankLine \tcp{Step 2: copula rescaling}
Re-sample $N$ degree triplets~$\k_j$ from the set of~$\k_i$;\\
\ForAll{$m=1,2,3$} {
Let $k_m$ be the list of the $m^\text{th}$ component values of vectors~$\k_j$;\\
Sort list $k_m$ in the non-degreasing order of values;\\
\ForAll{$j=1 \ldots N$} {
Let $r_m^j$ be the position number of value $k_m^j$ in the sorted list;\\
$u_m^j = r_m^j/N$. } }

\BlankLine \tcp{Step 3: merge rescaled ADs and the ADD copula}
\ForAll{$m=1,2,3$} { \ForAll{$j=1 \ldots N$} { $q_m^j =
D_m^{-1}(u_m^j)$. } }

\BlankLine \KwOut{Degree vectors $\mathbf{q}_j =
(q_1^j,q_2^j,q_3^j)$, \mbox{$j=1 \ldots N$}, of the synthetic
topology.}
\end{algorithm}

We conclude our discussion of rescaling with the following remark.
Recall from the end of Section~\ref{sec:as-annotations} that we have
the following two types of multivariate statistics: the ADD and the
JDDs. As illustrated in Figure~\ref{fig:road}, we rescale the ADD
using all the three steps described in this section. For rescaling a
JDD, it is not necessary to separately rescale its marginals, i.e.,
to use the first step of the described rescaling process, since the
marginals of JDDs are distributions of scalar degrees that we
automatically rescale during the ADD rescaling. To rescale a JDD, we
execute only the second step of the described rescaling process to
obtain the rescaled empirical JDD copula. We then use this copula to
determine proper placement of edges in the final synthetic graph
that we construct. In other words, the last, third step of our
multivariate rescaling process applied to JDDs takes place during
the graph construction phase, which we describe next.

\subsection{Construction}
\label{sec:relmod:gen:gen}

We describe the $1K$- and
$2K$-annotated random graph constructors that are both
generalizations of the well-known {\em configuration\/} or {\em
pseudograph\/} approach in the terminology of~\cite{MaKrFaVa06}. The
$1K$-constructor requires only the rescaled ADD, while the
$2K$-constructor needs also the rescaled JDD copulas.

\subsubsection{Constructing $1K$-annotated random graphs}

Using the rescaled degree vectors~$\mathbf{q}_j$, \mbox{$j=1 \ldots
N$}, we construct $1K$-annotated random graphs using the following
algorithm:
\begin{enumerate}
\item for each vector~$\mathbf{q}_j =(q^j_1,q^j_2,q^j_3)$, prepare a
node with $q^j_1$ customer stubs, $q^j_2$ provider stubs, and $q^j_3$
peer stubs;
\item randomly select pairs of either customer-and-provider or
peer-and-peer stubs, and connect (match) them together to form c2p
or p2p links;
\item remove unmatched stubs, multiple edges between
the same pair of nodes (loops), links with both ends connected to
the same node (self-loops), and extract the largest connected
component.
\end{enumerate}

The last step deals with the known problem of the pseudograph
approach. As its name suggests, it does not necessarily produce
simple connected graphs. In general, it generates pseudographs,
i.e., graphs with (self-)loops, consisting of several connected
components. The size of the largest connected component is usually
comparable with the total pseudograph size, while all others are
small. Extraction of this largest connected component and removal of
all (self-)loops\footnote{Self-loops are removed, while multiple
edges between the same pair of nodes are mapped to a single edge
between the two nodes.} alters the target degree distributions.
Therefore, the resulting simple connected graph has a slightly
different ADD than the one on the algorithm input.

Annotations alleviate this problem since they introduce a series of
additional constraints. For example, in the non-annotated case,
loops tend to form between high-degree ASs, simply because these ASs
have a lot of stubs attached to them after step~1 of the algorithm.
In the annotated case, the number of such loops is smaller because
most stubs attached to high-degree ASs are annotated as provider
stubs that can be matched only with customer stubs attached mostly
to low-degree ASs.

Still, the $1K$-annotated random graphs are not perfect as, for
example, p2p links might end up connecting nodes with drastically
dissimilar degree, cf.~the JDD discussion in
Section~\ref{sec:as-annotations}. The $2K$-annotated random graphs
do not have this problem.

\subsubsection{Constructing $2K$-annotated random graphs}

Earlier work~\cite{MaKrFaVa06} extends the pseudograph approach to
non-annotated $2K$-distributions. We extend it even further for the
$2K$-annotated case in the following algorithm:
\begin{enumerate}

\item for each vector~$\mathbf{q}_j =(q^j_1,q^j_2,q^j_3)$, prepare a
node with $q^j_1$ customer stubs, $q^j_2$ provider stubs, $q^j_3$ peer
stubs, and total degree~$q^j = |\mathbf{q}_j|_1$;

\item determine the
total numbers $n_{c2p}$ and $n_{p2p}$ of c2p and p2p edges in the
target graph as the maximum possible number of customer-and-provider
and peer-and-peer stubs that can be matched within the stub
collection~$\mathbf{q}_j$;

\item rescale the c2p and p2p JDD copulas\footnote{See the remark at the end of
Section~\ref{sec:rescaling-multivariate}.}
to target sizes of~$n_{c2p}$ and $n_{p2p}$ degree pairs~$(q,q')$
corresponding to c2p and p2p edges between nodes of total
degrees~$q$ and~$q'$ in the target graph;

\item for each c2p (or p2p) degree pair~$(q,q')$ select randomly a
customer (or peer) stub attached to a node of degree~$q$ and a
provider (or peer) stub attached to a node of degree~$q'$ and form a
c2p (or p2p) edge;

\item use the procedure described below to rewire \mbox{(self-)}loops;

\item remove unmatched stubs, remaining \mbox{(self-)}loops,
 and extract the largest connected component.

\end{enumerate}

The following rewiring procedure reduces the number of edges removed
from the final graph. For each edge involved in a \mbox{(self-)}loop
between nodes of degrees~$q_1$ and~$q_2$, we randomly select two
non-adjacent nodes of degrees~$q_1$ and~$q_2$ and move the edge to
these nodes. This procedure retains a large number of edges that
would, otherwise, be removed from the graph. In theory, this
procedure may skew the original $2K$-summary statistics. In
practice, however, it alters these statistics negligibly.

The resulting graph has both the ADD and JDDs approximately the same
as those obtained after rescaling. Minor discrepancies are due to
the last step of the algorithm, but the number of
\mbox{(self-)}loops and small connected components are even smaller
than in the $1K$-annotated case. The reason for these improvements
is yet additional structural constraints, compared with the
$1K$-annotated case. For example, the JDD-induced constraints force
the algorithm to create only one link between a pair of high-degree
nodes, or no links between a pair of nodes of degree~1, thus
avoiding creation of many connected components composed of such node
pairs. The original graph does not have such links, and the rescaled
JDDs preserve these structural properties, thus improving the
resulting graph quality.

\section{Evaluation}
\label{sec:eva}

In this section, we present results of evaluation of our
$2K$-annotated graph generation method. We also evaluated the
$1K$-annotated generator and found that, as expected, it produced less
accurate graphs with defects such as those mentioned in
Section~\ref{sec:as-annotations}, e.g., with p2p links connecting ASs
of dissimilar degrees, etc.

{\bf Experiments.} To evaluate the accuracy of our $2K$-annotated
generator, we want to compare graphs it produces with the measured
annotated Internet AS graph from Section~\ref{sec:extraction}.
To simplify comparisons, we select one, most representative graph
from a set of 50 random synthetic graphs. We select this most
representative graph as follows. We first look for a simple graph
metric that exhibits high variability across the generated graphs.
One such metric is the maximum degree.
The expected maximum degree in an $n$-node graph with a power-law
degree distribution \mbox{$P(k) \sim k^{-\gamma}$} is $k_{max}
\approx n^{1/(\gamma-1)}$ \cite{BoPaVe04}. Exponent $\gamma$ is
approximately $2.1$ for the Internet AS topology. This value of
$\gamma$ stays constant as the Internet grows, and it implies almost
linear scaling of the maximum degree since $1/(\gamma-1) \approx
0.9$, which is consistent with scaling of maximum degree in
historical Internet topologies \cite{PaSaVe04-book}. For these
reasons, our most representative graph is the one with its maximum
degree closest to its expected value, across all the generated
graphs.

In addition, we evaluate the variance of important graph metrics described below,
across ensembles of random graphs that we generate.
Studying the variance properties of a graph generation technique is
essential for estimating structural differences between equal-sized random graphs generated by
the model, and for gaining insight on how such differences affect
performance evaluation experiments. The variance properties of a graph
generation technique is associated with the following
tradeoff. On the one hand, variance should be small so that generated
graphs closely match the observed topology. On the other
hand, though, random graphs should not all be identical or almost
identical, because if they do not exhibit sufficient structural diversity,
then they have little value for performance evaluation
studies. In our experiments, we compute and report the variance of
important graph metrics in sets of 50 equal-sized random graphs.

{\bf Metrics.} Since it is practically impossible to compare graphs
over every existing graph metric, we select a set of metrics that
were found particularly important in the Internet topology
literature. These metrics include the degree distributions that
we deal with in previous sections, assortativity coefficient,
distance distribution, and spectrum.  The {\it assortativity
coefficient} is essentially the Pearson correlation coefficient of
the joint degree distribution~(JDD). Its positive (negative) values
indicate that degrees of connected nodes are positively (negatively)
correlated, meaning that nodes with similar (dissimilar) degrees
interconnect with higher probabilities. The {\it distance
distribution} is the distribution of lengths of the shortest paths
in a graph, which we compute both with and without constraints
imposed by annotations (routing policies). The {\it spectrum} of a
graph is the set of the eigenvalues of its Laplacian~$L$. The
Laplacian's matrix elements~$L_{ij}$ are $-1/(k_ik_j)^{1/2}$ if
there is an edge between node~$i$ of degree~$k_i$ and node~$j$ of
degree~$k_j$; 1 if~$i=j$; and 0 otherwise. Among the~$n$ eigenvalues
of~$L$, the smallest non-zero and largest eigenvalues are most
interesting, since they provide tight bounds to a number of
important network properties. For more details on these and other
metrics, and why they are important, see~\cite{MaKrFo06}.

\begin{figure}
\centering
\includegraphics[width=4in]{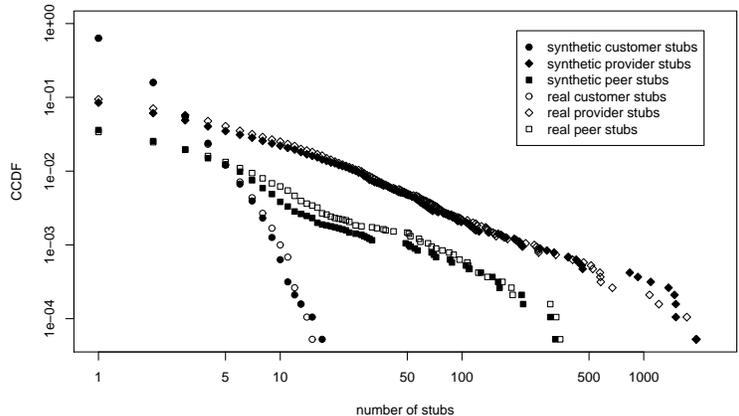}
\caption{CCDFs of the number of customer, provider, and peer stubs
in the synthetic versus measured AS topology.} \label{fig:ADs}
\end{figure}

{\bf Results.}  In Figure~\ref{fig:ADs} we plot the ADs of the
measured AS topology and of the most representative synthetic graph
of the equal size. We observe that the distributions of the
customer, provider, and peer degrees in the synthetic graph are very
close to the corresponding distributions in the measured topology.
The close match demonstrates that: 1)~spline-smoothing accurately
models complex ADs of real Internet topologies, 2)~random number
generation yields empirical distributions that follow the modeled
distributions, and 3)~rewiring and removal of \mbox{(self-)}loops do
not introduce any significant artifacts.
It is, of course, expected that our generator accurately reproduces
ADs, as they are part of the $2K$-summary statistics we explicitly
model. We also confirm that synthetic graphs, also as expected,
closely reproduce all the other summary statistics that we
explicitly model: the DD, ADD, and JDDs of the synthetic graph are
very close to the originals. We do not show the corresponding plots
for brevity.

\begin{figure}
\centering \subfigure[Shortest paths.]{
    \label{fig:spdist}
    \includegraphics[width=2.3in]{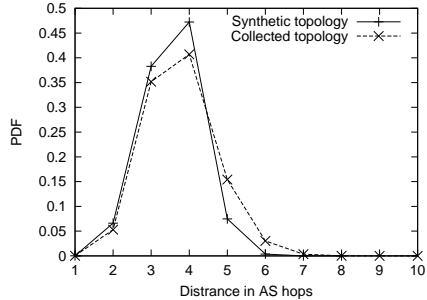}
    }
\subfigure[Shortest {\it valid} paths.]{
    \label{fig:vfdist}
    \includegraphics[width=2.3in]{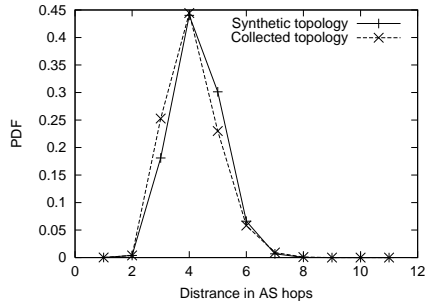}
    }
\caption{\label{fig:dist} Distance distributions
    in synthetic and measured AS topologies.}
\end{figure}

In Figures~\ref{fig:spdist} and~\ref{fig:vfdist} we compare the
distance distributions of the measured and equal-sized synthetic
topology ignoring and accounting for annotation-induced, i.e.,
routing policy, constraints. In the former case, we calculate
lengths of the standard shortest paths between nodes in a graph as
if the graph was non-annotated. In the latter case, we find lengths
of shortest {\em valid}, i.e., valley-free, paths defined in
Section~\ref{sec:short}. In both the non-annotated and annotated
cases, we observe that the distance distribution in the synthetic
graph closely matches the distance distribution in the measured
topology, even though we have {\em not\/} explicitly modeled or
tried to reproduce the distance distributions. We also observe that
the distance distributions in the non-annotated and annotated cases
are different, meaning that annotations in the synthetic graph
properly filter realistic, policy-constrained paths from the set of
all possible path in the non-annotated case.

In Table~\ref{table:metrics} we compare the measured topology with
synthetic graphs of different sizes over a set of important scalar
metrics, including those we do not explicitly model or try to
reproduce, e.g., the eigenvalues of the Laplacian, etc.
We compute these metrics for five synthetic graphs of sizes~$5,000$,
$10,000$, $30,000$, and $19,036$ nodes, the last size being equal to
the size of the original topology. The first three metrics are the
number of (c2p or p2p) edges in a graph. We observe that the number
of such edges grows almost linearly with the number of nodes. This
observation is consistent with that the average degree in historical
Internet topologies stays almost constant \cite{PaSaVe04-book}. The
fourth metric is the maximum degree. As expected, the maximum degree
grows with the size of the graph slightly slower than linearly. The
next five metrics in the table describe properties that have stayed
relatively constant in historical Internet topologies. These
properties have small variations in the synthetic graphs as well.

\begin{table*}
\centering \caption{\label{table:metrics} Scalar metrics of
synthetic and collected graphs. Note that smallest eigenvalues are
positive, but some may round to zero.} \resizebox{\textwidth}{!}{

\begin{tabular}{|c||c|c|c|c|c|}
\hline
& \multicolumn{5}{|c|}{Number of nodes} \\
\hline
\multirow{2}{*}{Metric}     & 19036  & 5000  & 10000 &  19036  & 30000 \\
                        & (measured) & (synthetic) & (synthetic) & (synthetic) & (synthetic) \\
\hline
Number of edges         & 40115 & 10179  & 20730   &  39595 &  62853   \\
Number of c2p edges         & 36188 & 9409   & 18917   &  36146 &  56870   \\
Number of p2p edges         & 3927  & 770    & 1813    &  3448  &  5983    \\
Maximum degree          & 2384  & 1014   & 1492    &  2385  &  3461    \\
Average degree          & 4.21  & 4.07   & 4.15    &  4.16  &  4.19    \\
Assortativity coefficient   & -0.20 & -0.30  & -0.24   & -0.25  &  -0.18   \\
Largest eigenvalue of Laplacian & 1.97  & 1.85   & 1.88    & 1.91   &  1.92    \\
Smallest eigenvalue of Laplacian & 0.03  &  0.00 & 0.12    & 0.09   &  0.00   \\
Average distance        & 3.76  & 3.26   & 3.52    & 3.57   &  3.75    \\

\hline
\end{tabular}
}
\end{table*}

\begin{figure*}
\centering
\subfigure[Total degree vs.\ number of customers.]{\label{fig:scatter:1}\includegraphics[width=.31\textwidth]{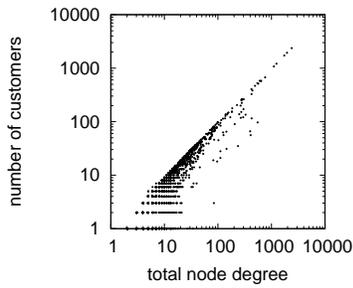}}\goodgap
\subfigure[Total degree vs.\ number of providers.]{\label{fig:scatter:2}\includegraphics[width=.31\textwidth]{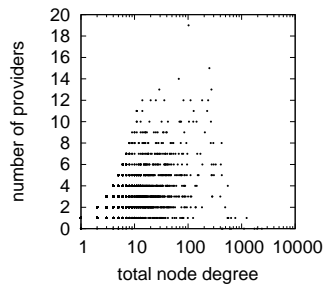}}\goodgap
\subfigure[Total degree vs.\ number of peers.]{\label{fig:scatter:3}\includegraphics[width=.31\textwidth]{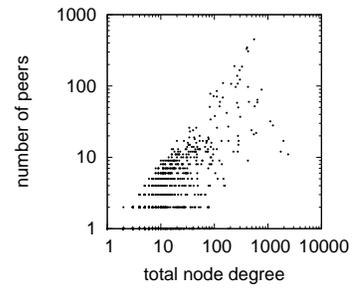}}%
\caption{\label{fig:scatter} Scatterplots demonstrating the diversity
  of annotation degrees and that total node degrees are agnostic with
  respect to annotations.}

\end{figure*}

Next we investigate the benefit of modeling the
ADs and ADD in addition to the DD and JDDs. Previous
work~\cite{MaKrFaVa06} shows that modeling DD and JDD is
sufficient for capturing and reliably reproducing most important
{\em non-annotated\/} graph metrics.
The main value of modeling ADs and ADD is that our generated synthetic
graphs are properly annotated. We saw in Section~\ref{sec:short} that the
Internet topology annotations are important. Here we provide another evidence
that they are non-trivial. Specifically,
in Figures~\ref{fig:scatter:1},~\ref{fig:scatter:2},
and~\ref{fig:scatter:3} we plot the total degrees of ASs in the
measured AS topology versus their annotation degrees: the number of customers~$k_1$,
providers~$k_2$, and peers~$k_3$, respectively. We observe that a
given total degree can correspond to a wide range of different values
of~$k_1$, $k_2$, and $k_3$. The JDD provides information only on total node degrees and
on their correlations, whereas it is completely agnostic to annotation
degrees. On the other hand, the ADs and ADD capture the distribution
of annotation degrees and the correlations between annotation degrees,
respectively. Therefore, the JDD alone is in principle incapable of
capturing topology annotations, while the benefit of modeling ADs and ADD lies in
reproducing realistic annotations in generated graphs.

To quantify the variance properties of randomly generated
graphs, we compute the standard deviation of our metrics across sets of 50
random graphs. We construct 4 sets with topologies of 5,000, 10,000,
19,036, and 30,000 nodes, a total of 200 random graphs. Among our
evaluation metrics, we do not compute the eigenvalues of the Laplacian
and the assortativity coefficient, since they require prohibitively
long computation times for 200 graphs. In
Table~\ref{table:var}, we show the standard deviation and mean
value of the remaining metrics. The maximum degree exhibits the
highest standard deviation (with respect to the mean) taking values
between 376 and 471 for graphs of different size. The high variance of
the maximum degree is expected, since the degree distribution of
Internet topologies is highly skewed. On the other hand, the remaining
metrics in Table~\ref{table:var} exhibit low variance. These metrics
reflect aggregate graph properties and can be modeled as a sum of many
i.i.d.\ random variables. Therefore, according to the central limit
theorem, their distribution is approximately normal and their variance
is consequently smaller than the variance of the maximum degree.

An important difference between the graph generation method described
in this study and the graph generation methods described in our
previous work~\cite{MaKrFaVa06} is that the former exhibits higher
variance. The two methods are conceptually similar in generating
synthetic graphs that reproduce the correlation profile
of an observed topology---albeit \cite{MaKrFaVa06} does not consider
annotations. They differ in that our previous techniques
directly use the degree distribution or correlations of an observed
topology to generate new similar topologies. On the other hand, the
present work first models the degree correlations of a topology and then
uses random number generators to produce synthetic degree distributions
fed into final graph constructors.
In simpler words, our present technique induces more randomness by means of
the synthetic generation of degree correlations based on the correlation
profile extracted from the real topology.
The two approaches are complementary and together
provide a wider range of options for generating synthetic topologies
with desired variance characteristics.

\begin{table}
\centering
\caption{\label{table:var} Variance of graph metrics across sets of equal-sized random graphs.}

\begin{tabular}{|c|c|c|c|c|}

\hline

\multirow{3}{*}{graph metric} & \multicolumn{4}{|c|}{ std.\ deviation / mean } \\
\cline{2-5}
                & 5000 & 10000 & 19036 & 30000 \\ & nodes & nodes &
                nodes & nodes \\ \hline

c2p edges   &  399/17,905  &   267/36,876  &  349/71,055  &  410/112,549   \\  \hline
p2p edges   &  100/1,238   &   203/2,980   &  396/6,412   &  541/12,863    \\  \hline
max degree      &  387/1,618   &   417/2,090   &  471/2,335   &  376/2,599     \\  \hline
av.\ degree      &  0.08/3.83   &   0.03/3.99   &  0.02/4.07   &  0.02/4.11     \\  \hline
av.\ distance    &  0.13/3.16   &   0.09/3.40   &  0.10/3.61   &  0.06/3.77     \\  \hline

\end{tabular}

\end{table}

Overall, our evaluation results show that:
\begin{itemize}
\item $2K$-annotated random graphs generated with our approach faithfully
reproduce a number of important properties of Internet topologies;

\item rescaled graphs exhibit the expected behavior
according to a number of definitive graph metrics, i.e., these
metrics are either properly-rescaled or stay relatively stable as
the size of synthetic graphs varies;

\item the profile of correlations between annotation
and total degrees is diverse; and

\item random graphs generated with our method
exhibit small variance, although higher than in our previous
work~\cite{MaKrFaVa06}.

\end{itemize}

\section{Conclusions}
\label{sec:con}

In this work, we have focused on the problem of generating synthetic
annotated graphs that model real complex networks. Our techniques are
likely to have many applications not only in networking, but also in
other disciplines where annotated graphs are used to abstract and represent
network structure. For example, two groups have recently contacted us
to discuss our techniques as they were searching for tools to generate
synthetic, semantic-rich, i.e., annotated, networks for their
simulation studies. The first group works on modeling the European
powerline networks, while the second is in brain and neural network
research. Other networks to which our techniques are immediately
applicable include the router-level Internet, WWW, networks of
critical resources dependencies, as well as many types of social
and biological networks, such as regulatory pathways~\cite{PaKo07}.

A number of open problems remain. In particular, our techniques
construct synthetic versions of real topologies available from
measurement projects. However, it is well-known that in many cases,
the outcome of measurements does not accurately represent a real
complete topology. In fact, there might exist inherent limitations
in measuring certain network topologies with 100\% accuracy. A venue
for further research is the development of prediction techniques
that extrapolate what we can presently measure in order to predict
what we can not measure.

Another substantial problem is the difficulties in validating
results of topology inference studies. For example, in the specific
context of Internet topologies, validation is hard because of the
unwillingness of service providers to release data on their
infrastructure, network design, configuration, and performance. On
the other hand, validation of any research result is a cornerstone
to its reliability and utility. Therefore we believe it is
imperative to focus on new validation techniques that would combine
the limited ground truth data available today with convincing
testbed or simulation experiments.

\section{Acknowledgments}

This work was supported in part by NSF CNS-0434996 and CNS-0722070,
by DHS N66001-08-C-2029, and by Cisco Systems.

\balance

\end{document}